\documentclass[]{JHEP}        
\usepackage{epsfig}

\def\la{\mathrel{\mathpalette\fun <}}
\def\ga{\mathrel{\mathpalette\fun >}}
\def\fun#1#2{\lower3.6pt\vbox{\baselineskip0pt\lineskip.9pt
        \ialign{$\mathsurround=0pt#1\hfill##\hfil$\crcr#2\crcr\sim\crcr}}}
\def\beq{\begin{equation}}
\def\eeq{\end{equation}}
\def\beqa{\begin{eqnarray}}
\def\eeqa{\end{eqnarray}}

\preprint{\ }

\title {Electron neutrino opacity in magnetised media }

\author{Esteban Roulet\\ Depto. de F\'\i sica Te\'orica, 
Universidad de Valencia,\\
 E-46100, Burjasot, Valencia, Spain \\
Email: \email{roulet@goya.ific.uv.es}}

\abstract{ We study the effects of strong magnetic fields ($B\ga
10^{13}$~G) in the cross section for $\nu_e n\to p e$ scattering in the
presence of a degenerate electron background. This can be relevant for
the $\nu_e$ propagation in the proto--neutron star stage after
supernovae collapse. We find that for field strengths $B\ga 
10^{16}$~G$(E_\nu/10$~MeV$)^2$ the $\nu_e$ opacity is sizeably
affected by the magnetic field and can lead to a shift in the location of
the electron neutrino sphere towards lower densities.
 We discuss the implications that this may have for scenarios proposed to
explain the observed pulsar velocities.}

\received{}
\accepted{}

\begin{document}

There has recently been a renewed interest in the study of the
neutrino propagation at very high densities ($\rho>10^{10}$g/cm$^3$)
and in the presence of strong 
magnetic fields ($B>10^{13}$~G), in connection with the proposed
explanations for the observed pulsar recoil velocities, which could
arise from an asymmetry in the neutrino emission during the first
seconds after the collapse of massive stars
\cite{ch84,do85,vi95,kuz96,be96,ku96,nu97,ho97}.  
The high densities reached in the proto--neutron star stage result in
the trapping of the neutrinos within the so--called neutrino spheres,
i.e. the surfaces from which the optical depth becomes of order unity.
The location of these surfaces depends on the neutrino flavour because of
the effects of the charged current interactions, which allow the
$\nu_e$  to escape only from larger radii
(mainly because of the large $\nu_e n\to pe$ cross
section) than the $\nu_{\mu,\tau}$ (and their antineutrinos),
while the $\bar\nu_e$--sphere radius is in between the
previous ones (due mainly to the $\bar\nu_e p\to ne^+$ interaction).

The fact that the radii of these neutrino spheres is flavour dependent
is at the basis of the mechanism suggested by Kusenko and Segr\`e
\cite{ku96} to account for the pulsar kicks, which is based on the
resonant conversion $\nu_e\to\nu_\tau$ just in the  region between
those two spheres, so that trapped $\nu_e$s oscillating to $\nu_\tau$s
become free to escape. As was noticed by those authors, the large
magnetic fields which may be present in the proto--neutron star (as
evidenced by the large $B$ fields, $B\sim 10^{12}$--$2\times
10^{13}$~G, observed in young pulsars), may be
responsible for a distortion of the resonance density for the MSW
conversion and can lead then to an asymmetry in the average energy of the
emitted $\nu_\tau$s. For this mechanism to be successful, 
field strengths $B\ga 10^{14}$~G   would be required, and some
scenarios have been proposed in which such large fields 
 may be achievable at the epoch of the supernova collapse
 \cite{mu79,th92,bi92,bo95}, with the maximum conceivable strength
 being $B\sim
10^{18}$~G, a value beyond which the magnetic energy becomes larger
than the gravitational one.

We want here to study whether these large magnetic fields
 may actually affect
the $\nu_e n\to pe$ cross section which determines the $\nu_e$ opacity
itself and hence shift the location of the $\nu_e$--sphere. 
This can happen due to the modification of the phase space
distribution of the final state electrons in the presence of the
magnetic field. It has indeed been shown that the neutron decay and
electron capture processes \cite{ma69,fa69,la91} are affected significantly for
$B\ga B_c\equiv m_e^2/e=4.141\times 10^{13}$~G, in which case the spacing
between the first 
Landau levels becomes larger than the electron mass $m_e$. For
larger $B$ fields, the separation between Landau levels may become
comparable to the typical energies of the neutrinos emitted 
in supernovae, $E_{\nu_e}\sim
1$--30~MeV, and hence the neutrino opacity may also be changed. 

The reaction rates for the $\nu_e n$ interaction in the presence of
magnetic fields were considered in \cite{ch93}, focussing in the
conditions for $\beta$ equilibrium. Furthermore, 
the possible $B$ dependence of the $\nu_e n\to pe$ cross section has
 been recently invoked as an explanation of the pulsar kicks if an
asymmetry in the magnetic field distribution develops during the first
second after the collapse \cite{bi97}. A detailed evaluation of the
magnetic field effects on the $\nu_e$ opacity is however lacking and
will be the subject of the present work.

We will then concentrate in the study of the $\nu_e$ produced during
the Kelvin Helmholtz cooling phase~\cite{bu86,su94}, 
just after the time when the
shock produced by the core bounce dissociates the heavy elements
initially present and make the $\nu_e$ opacity in the shock heated
region to be dominated by the
reaction $\nu_en\to pe$ (before that, scattering off nuclei dominates).

At zero density and in the absence of magnetic fields, the total cross
section for this reaction is (see e.g. \cite{br85})
\beq
\sigma_0={G^2\over
\pi}(g_V^2+3g_A^2)(E_\nu+Q)^2\sqrt{1-m_e^2/(E_\nu+Q)^2},
\label{sigma0.eq}
\eeq
where $G\equiv G_F\cos\theta_c$ stands for the product of the Fermi
coupling and the Cabibbo mixing while $g_V=1$ and $g_A=1.26$ are the
vector and axial nucleon couplings. Since the energy transfered to the
recoiling proton is ${\cal O}(E_\nu^2/m_p)$, which is negligible for the 
neutrino energies we are interested in, one has $E_e\simeq E_\nu+Q$, with
$Q\equiv m_p-m_n=1.293$~MeV. This `elastic' approximation is indeed
quite reliable \cite{re97} 
for densities below the nuclear saturation one and for
non--degenerate neutrinos ($\mu_\nu\simeq 0$),  as is the case of
interest here for the study of the location of the neutrino--sphere. 

At finite densities (but still with $B=0$), the main modification will
come from an overall electron blocking factor $1-f_e(E_e)$ multiplying
the cross section, where
\beq
f_e(E)={1\over 1+{\rm exp}\left[(E-\mu)/T\right]},
\eeq
is the Fermi Dirac distribution at temperature
$T$, with $\mu$ being the electron's chemical potential.

Considering now a non--vanishing magnetic field, the matrix element
for the process will remain essentially unaffected (we average over
initial spins neglecting the neutron polarisation, which is small for
$B<10^{17}$~G) 
\cite{fa69,la91}.  The
main modification will then come from the available phase space for
the electrons, since the phase space factor for $B=0$
\beq
\sum_e=2\int {d^3p\over (2\pi)^3}
\eeq
has now to be replaced with
\beq
\sum_e={eB\over (2\pi)^2}\sum_{n=0}^{n_{max}}g_n\int dp_z,
\eeq
where $g_0=1$ and $g_n=2$ for $n\geq 1$ are the
degeneracies of the Landau levels of energy
\beq
E_n=\sqrt{p_z^2+m_e^2(1+2nB_*)},
\eeq
and we have introduced $B_*\equiv B/B_c$.

Computing the $\nu_en\to pe$ cross section in the presence of the
magnetic field we then obtain
\beqa
\sigma=&{G^2\over 4\pi}(g_V^2+3g_A^2)m_e^2B_*\sum_ng_n\int
dp_z\delta(E_\nu+Q-E_e) \left[1-f_e(E_e)\right]\nonumber \\ 
=&{G^2\over 2\pi}(g_V^2+3g_A^2)m_e^2B_*\left[1-f_e(E_\nu+Q)\right] 
(E_\nu+Q)\sum_ng_n/p^{(n)}_z,\hfill
\label{sigma.eq}
\eeqa
where $p^{(n)}_z\equiv \sqrt{(E_\nu+Q)^2-m_e^2(1+2nB_*)}$.
The maximum Landau level accessible to the final state electron,
for a given energy of the initial neutrino, is
\beq
n_{max}={\rm int}\left\{{1\over 2B_*}\left[\left(E_\nu+Q\over
m_e\right)^2-1 \right]\right\}.
\eeq

In figure~1  we plot the cross section in vacuum, normalised to the
$B=0$ one ($\Sigma\equiv \sigma/\sigma_0$), as a function of the
magnetic field, fixing $E_\nu=10$~MeV for definiteness.

\FIGURE{
\setlength{\unitlength}{0.240900pt}
\ifx\plotpoint\undefined\newsavebox{\plotpoint}\fi
\sbox{\plotpoint}{\rule[-0.500pt]{1.000pt}{1.000pt}}%

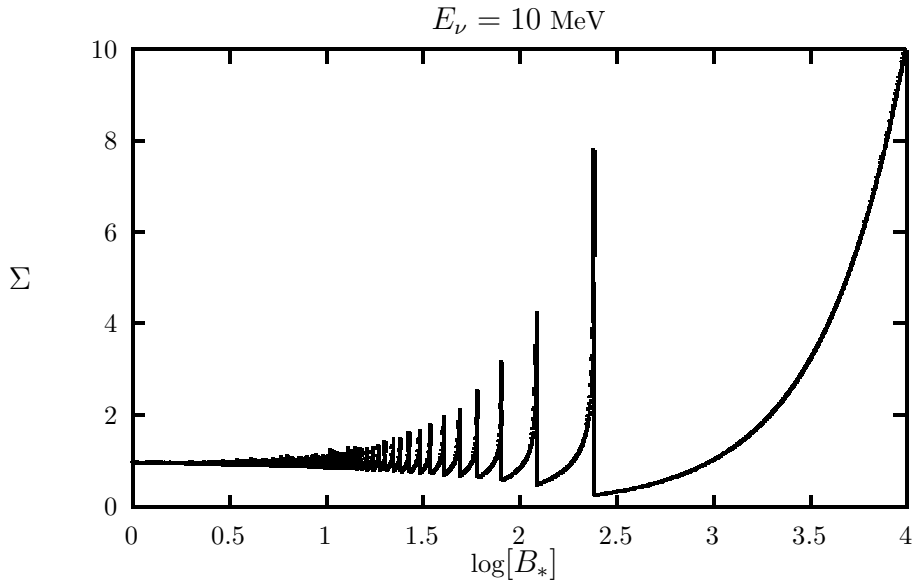
 \caption{Normalised cross section,
$\Sigma\equiv \sigma/\sigma_0$, for $E_\nu=10$~MeV, as a function of
the magnetic field $B_*\equiv B/B_c$, in the absence of background
matter and for $T=0$.}}

For field values 
\beq
B>{1\over 2}\left[\left({E_\nu+Q\over m_e}\right)^2-1\right]B_c\simeq
10^{16}{\rm G}
\left({E_\nu\over 10\ {\rm MeV}}\right)^2,
\eeq
only the lowest Landau level ($n=0$) contributes to the phase space,
and in this case one has
\beq
\sigma\simeq {G^2\over 2\pi}(g_V^2+3g_A^2)m_e^2B_*,
\eeq
which grows linearly with $B$ and is independent of the neutrino
energy. For smaller magnetic field values, $n_{max}\geq 1$ and hence 
more Landau levels
contribute to the sum in eq.~(\ref{sigma.eq}). The singular behaviour
present each time that a new Landau level opens up, which is similar
to the one found in the $\beta$--spectrum in $n$--decay \cite{fa69},  
arises from the
$p_z^{-1}$ factor from $dp_z=E/p_zdE$, and is expected to be somewhat
smeared once the proton recoil momentum is included and its effects also
averaged out once a distribution of neutrino energies is considered. 
 In the limit of small magnetic fields, $n_{max}$ will be large and we
may approximate the sum over Landau levels as
\beqa
\sum_{n=0}^{n_{max}}{g_n\over p^{(n)}_z}=&(E_e^2-m_e^2)^{-1/2}+2 
\sum_{n=1}^{n_{max}}(E_e^2-m_e^2(1+2nB_*))^{-1/2} \hfill\nonumber\\
\simeq & (E_e^2-m_e^2)^{-1/2}\left[1+{x\over B_*}
\int_{2B_*/x}^1 dy(1-y)^{-1/2}\right] \hfill \\=&
(E_e^2-m_e^2)^{-1/2}\left[1+{2x\over
B_*}\sqrt{1-2B_*/x}\right],\hfill\nonumber 
\eeqa
where we defined $x\equiv (E_e/m_e)^2-1$. From this we get, for
$B_*\ll x$,
\beq
\Sigma\simeq 1-{B_*\over 2x},
\eeq
and the $B=0$ result in eq.~(\ref{sigma0.eq}) 
is then asymptotically recovered.

In the presence of background matter, 
the electron density is given by
\beq
n_e=\sum_ef_e(E)={m_e^2B_*\over (2\pi)^2}\sum_ng_n\int dp_zf_e(E_n),
\eeq
and a similar expression  holds for the positron density, with the
replacement $\mu\to -\mu$.

Taking into account that for typical proto--neutron star temperatures
a positron background is also generally present, we can introduce the
electron fraction as
\beq
Y_e={n_e-n_{e^+}\over n_p+n_n}.
\eeq
The star density may then be written as
\beq
\rho\simeq m_p(n_p+n_n)={m_p\over Y_e}(n_e-n_{e^+}).
\eeq

\FIGURE{
\setlength{\unitlength}{0.240900pt}
\ifx\plotpoint\undefined\newsavebox{\plotpoint}\fi
\sbox{\plotpoint}{\rule[-0.500pt]{1.000pt}{1.000pt}}%
\begin{picture}(1500,900)(0,0)
\font\gnuplot=cmr10 at 10pt
\gnuplot
\sbox{\plotpoint}{\rule[-0.500pt]{1.000pt}{1.000pt}}%
\put(220.0,113.0){\rule[-0.500pt]{292.934pt}{1.000pt}}
\put(220.0,113.0){\rule[-0.500pt]{4.818pt}{1.000pt}}
\put(198,113){\makebox(0,0)[r]{0}}
\put(1416.0,113.0){\rule[-0.500pt]{4.818pt}{1.000pt}}
\put(220.0,233.0){\rule[-0.500pt]{4.818pt}{1.000pt}}
\put(198,233){\makebox(0,0)[r]{2}}
\put(1416.0,233.0){\rule[-0.500pt]{4.818pt}{1.000pt}}
\put(220.0,353.0){\rule[-0.500pt]{4.818pt}{1.000pt}}
\put(198,353){\makebox(0,0)[r]{4}}
\put(1416.0,353.0){\rule[-0.500pt]{4.818pt}{1.000pt}}
\put(220.0,473.0){\rule[-0.500pt]{4.818pt}{1.000pt}}
\put(198,473){\makebox(0,0)[r]{6}}
\put(1416.0,473.0){\rule[-0.500pt]{4.818pt}{1.000pt}}
\put(220.0,592.0){\rule[-0.500pt]{4.818pt}{1.000pt}}
\put(198,592){\makebox(0,0)[r]{8}}
\put(1416.0,592.0){\rule[-0.500pt]{4.818pt}{1.000pt}}
\put(220.0,712.0){\rule[-0.500pt]{4.818pt}{1.000pt}}
\put(198,712){\makebox(0,0)[r]{10}}
\put(1416.0,712.0){\rule[-0.500pt]{4.818pt}{1.000pt}}
\put(220.0,832.0){\rule[-0.500pt]{4.818pt}{1.000pt}}
\put(198,832){\makebox(0,0)[r]{12}}
\put(1416.0,832.0){\rule[-0.500pt]{4.818pt}{1.000pt}}
\put(220.0,113.0){\rule[-0.500pt]{1.000pt}{4.818pt}}
\put(220,68){\makebox(0,0){9}}
\put(220.0,812.0){\rule[-0.500pt]{1.000pt}{4.818pt}}
\put(372.0,113.0){\rule[-0.500pt]{1.000pt}{4.818pt}}
\put(372,68){\makebox(0,0){9.5}}
\put(372.0,812.0){\rule[-0.500pt]{1.000pt}{4.818pt}}
\put(524.0,113.0){\rule[-0.500pt]{1.000pt}{4.818pt}}
\put(524,68){\makebox(0,0){10}}
\put(524.0,812.0){\rule[-0.500pt]{1.000pt}{4.818pt}}
\put(676.0,113.0){\rule[-0.500pt]{1.000pt}{4.818pt}}
\put(676,68){\makebox(0,0){10.5}}
\put(676.0,812.0){\rule[-0.500pt]{1.000pt}{4.818pt}}
\put(828.0,113.0){\rule[-0.500pt]{1.000pt}{4.818pt}}
\put(828,68){\makebox(0,0){11}}
\put(828.0,812.0){\rule[-0.500pt]{1.000pt}{4.818pt}}
\put(980.0,113.0){\rule[-0.500pt]{1.000pt}{4.818pt}}
\put(980,68){\makebox(0,0){11.5}}
\put(980.0,812.0){\rule[-0.500pt]{1.000pt}{4.818pt}}
\put(1132.0,113.0){\rule[-0.500pt]{1.000pt}{4.818pt}}
\put(1132,68){\makebox(0,0){12}}
\put(1132.0,812.0){\rule[-0.500pt]{1.000pt}{4.818pt}}
\put(1284.0,113.0){\rule[-0.500pt]{1.000pt}{4.818pt}}
\put(1284,68){\makebox(0,0){12.5}}
\put(1284.0,812.0){\rule[-0.500pt]{1.000pt}{4.818pt}}
\put(1436.0,113.0){\rule[-0.500pt]{1.000pt}{4.818pt}}
\put(1436,68){\makebox(0,0){13}}
\put(1436.0,812.0){\rule[-0.500pt]{1.000pt}{4.818pt}}
\put(220.0,113.0){\rule[-0.500pt]{292.934pt}{1.000pt}}
\put(1436.0,113.0){\rule[-0.500pt]{1.000pt}{173.207pt}}
\put(220.0,832.0){\rule[-0.500pt]{292.934pt}{1.000pt}}
\put(45,472){\makebox(0,0){$\Sigma$}}
\put(828,23){\makebox(0,0){log$[\rho (Y_e/0.1)]$ (g/cm$^3$) }}
\put(828,877){\makebox(0,0){$E_\nu=10$ MeV}}
\put(342,143){\makebox(0,0)[l]{$B_*=1.$E2}}
\put(828,203){\makebox(0,0)[l]{$B_*=1.$E3}}
\put(1132,760){\makebox(0,0)[l]{$B_*=1.$E4}}
\put(220.0,113.0){\rule[-0.500pt]{1.000pt}{173.207pt}}
\sbox{\plotpoint}{\rule[-0.300pt]{0.600pt}{0.600pt}}%
\multiput(820.99,137.97)(0.502,-6.243){5}{\rule{0.121pt}{6.270pt}}
\multiput(818.75,150.99)(5.000,-37.986){2}{\rule{0.600pt}{3.135pt}}
\put(220.0,164.0){\rule[-0.300pt]{144.540pt}{0.600pt}}
\put(825.0,113.0){\rule[-0.300pt]{134.663pt}{0.600pt}}
\sbox{\plotpoint}{\rule[-0.250pt]{0.500pt}{0.500pt}}%
\multiput(220,174)(12.453,0.000){30}{\usebox{\plotpoint}}
\put(593.60,174.00){\usebox{\plotpoint}}
\multiput(603,174)(12.453,0.000){2}{\usebox{\plotpoint}}
\put(630.96,174.00){\usebox{\plotpoint}}
\multiput(640,174)(12.453,0.000){2}{\usebox{\plotpoint}}
\put(668.32,174.00){\usebox{\plotpoint}}
\put(680.77,174.00){\usebox{\plotpoint}}
\put(693.23,174.00){\usebox{\plotpoint}}
\put(705.68,174.00){\usebox{\plotpoint}}
\put(718.13,174.00){\usebox{\plotpoint}}
\put(730.59,174.00){\usebox{\plotpoint}}
\put(743.04,174.00){\usebox{\plotpoint}}
\put(755.49,174.00){\usebox{\plotpoint}}
\put(767.95,174.00){\usebox{\plotpoint}}
\put(780.40,174.00){\usebox{\plotpoint}}
\put(792.85,174.00){\usebox{\plotpoint}}
\put(805.31,174.00){\usebox{\plotpoint}}
\put(817.76,174.00){\usebox{\plotpoint}}
\put(830.21,174.00){\usebox{\plotpoint}}
\put(842.67,174.00){\usebox{\plotpoint}}
\put(855.12,174.00){\usebox{\plotpoint}}
\put(867.57,174.00){\usebox{\plotpoint}}
\put(880.03,174.00){\usebox{\plotpoint}}
\put(892.48,174.00){\usebox{\plotpoint}}
\put(904.93,174.00){\usebox{\plotpoint}}
\put(917.39,174.00){\usebox{\plotpoint}}
\put(929.84,174.00){\usebox{\plotpoint}}
\put(942.29,174.00){\usebox{\plotpoint}}
\put(954.75,174.00){\usebox{\plotpoint}}
\put(967.20,174.00){\usebox{\plotpoint}}
\multiput(977,174)(2.404,-12.219){5}{\usebox{\plotpoint}}
\put(991.75,113.00){\usebox{\plotpoint}}
\put(1004.20,113.00){\usebox{\plotpoint}}
\put(1016.66,113.00){\usebox{\plotpoint}}
\put(1029.11,113.00){\usebox{\plotpoint}}
\put(1041.56,113.00){\usebox{\plotpoint}}
\put(1054.02,113.00){\usebox{\plotpoint}}
\put(1066.47,113.00){\usebox{\plotpoint}}
\put(1078.92,113.00){\usebox{\plotpoint}}
\put(1091.38,113.00){\usebox{\plotpoint}}
\put(1103.83,113.00){\usebox{\plotpoint}}
\put(1116.28,113.00){\usebox{\plotpoint}}
\put(1128.73,113.00){\usebox{\plotpoint}}
\put(1141.19,113.00){\usebox{\plotpoint}}
\put(1153.64,113.00){\usebox{\plotpoint}}
\put(1166.09,113.00){\usebox{\plotpoint}}
\put(1178.55,113.00){\usebox{\plotpoint}}
\put(1191.00,113.00){\usebox{\plotpoint}}
\put(1203.45,113.00){\usebox{\plotpoint}}
\put(1215.91,113.00){\usebox{\plotpoint}}
\put(1228.36,113.00){\usebox{\plotpoint}}
\put(1240.81,113.00){\usebox{\plotpoint}}
\put(1253.27,113.00){\usebox{\plotpoint}}
\put(1265.72,113.00){\usebox{\plotpoint}}
\put(1278.17,113.00){\usebox{\plotpoint}}
\put(1290.63,113.00){\usebox{\plotpoint}}
\put(1303.08,113.00){\usebox{\plotpoint}}
\put(1315.53,113.00){\usebox{\plotpoint}}
\put(1327.99,113.00){\usebox{\plotpoint}}
\put(1340.44,113.00){\usebox{\plotpoint}}
\put(1352.89,113.00){\usebox{\plotpoint}}
\put(1365.35,113.00){\usebox{\plotpoint}}
\put(1377.80,113.00){\usebox{\plotpoint}}
\put(1390.25,113.00){\usebox{\plotpoint}}
\put(1402.71,113.00){\usebox{\plotpoint}}
\multiput(1403,113)(12.453,0.000){0}{\usebox{\plotpoint}}
\put(1415.16,113.00){\usebox{\plotpoint}}
\put(1427.61,113.00){\usebox{\plotpoint}}
\put(1436,113){\usebox{\plotpoint}}
\multiput(220,728)(29.058,0.000){21}{\usebox{\plotpoint}}
\put(830.21,728.00){\usebox{\plotpoint}}
\put(859.27,728.00){\usebox{\plotpoint}}
\multiput(860,728)(29.058,0.000){0}{\usebox{\plotpoint}}
\put(888.33,728.00){\usebox{\plotpoint}}
\multiput(897,728)(29.058,0.000){0}{\usebox{\plotpoint}}
\put(917.38,728.00){\usebox{\plotpoint}}
\multiput(926,728)(29.058,0.000){0}{\usebox{\plotpoint}}
\put(946.44,728.00){\usebox{\plotpoint}}
\multiput(952,728)(29.058,0.000){0}{\usebox{\plotpoint}}
\multiput(964,728)(29.058,0.000){0}{\usebox{\plotpoint}}
\put(975.50,728.00){\usebox{\plotpoint}}
\multiput(986,728)(29.058,0.000){0}{\usebox{\plotpoint}}
\put(1004.56,728.00){\usebox{\plotpoint}}
\multiput(1008,728)(29.058,0.000){0}{\usebox{\plotpoint}}
\multiput(1018,728)(29.058,0.000){0}{\usebox{\plotpoint}}
\put(1033.62,728.00){\usebox{\plotpoint}}
\multiput(1038,728)(29.058,0.000){0}{\usebox{\plotpoint}}
\multiput(1048,728)(29.058,0.000){0}{\usebox{\plotpoint}}
\put(1062.67,728.00){\usebox{\plotpoint}}
\multiput(1068,728)(29.058,0.000){0}{\usebox{\plotpoint}}
\multiput(1077,728)(29.058,0.000){0}{\usebox{\plotpoint}}
\put(1091.73,728.00){\usebox{\plotpoint}}
\multiput(1096,728)(29.058,0.000){0}{\usebox{\plotpoint}}
\multiput(1106,728)(29.058,0.000){0}{\usebox{\plotpoint}}
\put(1120.79,728.00){\usebox{\plotpoint}}
\multiput(1125,728)(29.058,0.000){0}{\usebox{\plotpoint}}
\multiput(1134,728)(29.058,0.000){0}{\usebox{\plotpoint}}
\put(1149.85,728.00){\usebox{\plotpoint}}
\multiput(1152,728)(29.058,0.000){0}{\usebox{\plotpoint}}
\multiput(1162,728)(29.058,0.000){0}{\usebox{\plotpoint}}
\put(1178.90,728.00){\usebox{\plotpoint}}
\multiput(1180,728)(29.058,0.000){0}{\usebox{\plotpoint}}
\multiput(1189,728)(29.058,0.000){0}{\usebox{\plotpoint}}
\put(1207.96,728.00){\usebox{\plotpoint}}
\multiput(1208,728)(29.058,0.000){0}{\usebox{\plotpoint}}
\multiput(1217,728)(29.058,0.000){0}{\usebox{\plotpoint}}
\multiput(1226,728)(29.058,0.000){0}{\usebox{\plotpoint}}
\put(1237.02,728.00){\usebox{\plotpoint}}
\multiput(1244,728)(29.058,0.000){0}{\usebox{\plotpoint}}
\multiput(1254,728)(29.058,0.000){0}{\usebox{\plotpoint}}
\put(1266.08,728.00){\usebox{\plotpoint}}
\multiput(1272,728)(29.058,0.000){0}{\usebox{\plotpoint}}
\multiput(1281,728)(0.425,-29.055){21}{\usebox{\plotpoint}}
\multiput(1290,113)(29.058,0.000){0}{\usebox{\plotpoint}}
\put(1299.25,113.00){\usebox{\plotpoint}}
\multiput(1308,113)(29.058,0.000){0}{\usebox{\plotpoint}}
\multiput(1318,113)(29.058,0.000){0}{\usebox{\plotpoint}}
\put(1328.31,113.00){\usebox{\plotpoint}}
\multiput(1336,113)(29.058,0.000){0}{\usebox{\plotpoint}}
\multiput(1345,113)(29.058,0.000){0}{\usebox{\plotpoint}}
\put(1357.37,113.00){\usebox{\plotpoint}}
\multiput(1363,113)(29.058,0.000){0}{\usebox{\plotpoint}}
\multiput(1372,113)(29.058,0.000){0}{\usebox{\plotpoint}}
\put(1386.43,113.00){\usebox{\plotpoint}}
\multiput(1391,113)(29.058,0.000){0}{\usebox{\plotpoint}}
\multiput(1400,113)(29.058,0.000){0}{\usebox{\plotpoint}}
\put(1415.48,113.00){\usebox{\plotpoint}}
\multiput(1418,113)(29.058,0.000){0}{\usebox{\plotpoint}}
\multiput(1427,113)(29.058,0.000){0}{\usebox{\plotpoint}}
\put(1436,113){\usebox{\plotpoint}}
\end{picture}
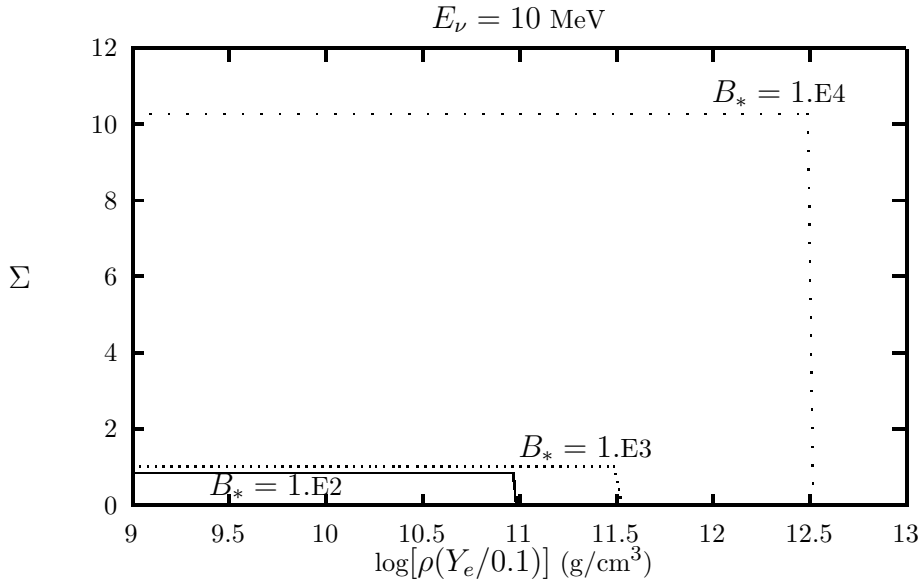
\caption{Normalised cross section vs.
background matter density, fixing $E_\nu=10$~MeV and for magnetic
fields $B_*=10^2,\ 10^3$ and $10^4$. For $B_*<10^2$ the results are
almost insensitive to $B$.}}

Considering now the effects of electron degeneracy in the $\nu_e n$
scattering,  we plot in
figure~2 the normalised cross section as a function of the density,
neglecting for the moment 
temperature effects ($T\ll E_\nu,\mu$), for different
values of $B$. Under this approximation $f_e(E)=\theta(\mu-E_\nu-Q)$,
and hence the effect of the background is just to block the final
state electrons, resulting in a maximum density $\rho_{max}$ 
beyond which no
scattering can take place and $\sigma$ vanishes. For densities larger
than $\rho_{max}$ only neutral
currents will contribute to the neutrino opacity\footnote{Here it 
is useful to recall that for $B=0,\ T=0$,
one has $\mu\simeq (3\pi^2 n_e)^{1/3}\simeq 
11\ {\rm MeV}(\rho Y_e/10^{10}$g/cm$^3)^{1/3}$, and
hence in this case $\rho_{max}Y_e\simeq
10^{10}$g/cm$^3((E_\nu+Q)/11$~MeV)$^3$.}. 
Increasing the size of $B$ will modify the size of $\sigma$ as
discussed in relation with figure~1, and, due to the $B$ dependence of
the electron density,  also change the maximum
density beyond which $\sigma$ vanishes, i.e.
$\rho_{max}=\rho_{max}(B,E_\nu)$.  One has to keep in mind that
large  magnetic fields will also affect the $Y_e$ values corresponding
to $\beta$ equilibrium and hence the details of the proto--neutron
star evolution.

\FIGURE{
\setlength{\unitlength}{0.240900pt}
\ifx\plotpoint\undefined\newsavebox{\plotpoint}\fi
\sbox{\plotpoint}{\rule[-0.500pt]{1.000pt}{1.000pt}}%

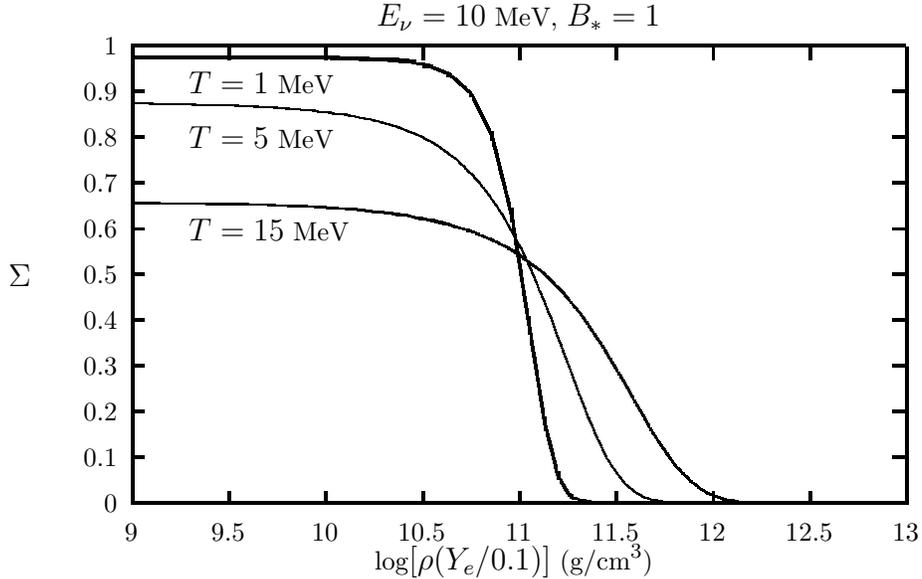
\caption{Temperature dependence of the
$\Sigma$ vs. $\rho$ relation, fixing $E_\nu=10$~MeV and $B_*=1$.}}

Turning now to the effects of finite temperatures, we show in figure~3
the modification of the previous picture with the inclusion of thermal
distributions, plotting $\Sigma$ vs. $\rho$ for different
temperatures. The main result is that some of the final states with
$E_e>\mu$ become occupied at finite temperatures, leading to a partial
Pauli blocking of the final state electrons, while for $E_e<\mu$,
 some electron states below the Fermi energy become free at
non--zero temperature and the available phase space does not vanish
abruptly but instead diminishes smoothly for increasing densities.
It is clear from figure~3 that only the
 low energy part of the neutrino spectrum is affected by
thermal effects, which just open up some phase space for the final
state electrons and result in a non--vanishing contribution to the
opacity up to much larger densities.

Let us now discuss the implications of the previous results for the
emission of neutrinos. During the cooling phase the neutrinos will
diffuse out from the inner regions of the proto--neutron
star\footnote{Actually, it is the lepton number which diffuses, since neutrinos
can be captured by neutrons and then reemited by electron captures, as
well as being thermally pair produced.} and
will be emitted from the surface of the neutrino sphere. The presence
of large magnetic fields can affect the electron neutrino opacity and
hence shift the radius of the neutrino sphere. A change in this radius
affects the neutrino luminosity in two ways: $i)$ it modifies the area of
the emitting surface ($L_\nu\propto R^2$) and $ii)$ the temperature of
the matter in the emission region can be different (the
temperature associated to the neutrinos is usually obtained from a fit
to the resulting neutrino spectrum, but the low energy neutrinos come
from deeper layers of the star, while those of higher energies are
trapped up to larger radii, making the neutrino spectrum not exactly a
Fermi Dirac one).

 Usually a mean neutrino
sphere radius  is obtained by averaging over the neutrino energy
distribution, but we prefer instead to work here with an energy dependent
neutrino sphere radius, since for instance when considering the
Kusenko and Segr\`e mechanism one needs to actually compare this radius 
 with the location of the
resonance for neutrino conversion, which is energy dependent.  

An estimate of the matter density at the $\nu_e$--sphere radius $R_e$
can be obtained from the condition of having unit optical depth,
$\tau(R_e)=1$, with
\beq
\tau(r)\equiv \int_r^\infty dr\ \sum_i n_i\sigma(\nu_e i\to X)\simeq
\sigma\int_r^\infty dr\ n_n(r),
\eeq
and where we used that the sum over possible scatterers $i$ is dominated
by the scattering $\nu_e n\to pe$.

Hence, we can write the approximate relation
\beq
1=\sigma\int_{R_e}^\infty drn_n(r)\simeq \sigma n_n(R_e) h_n,
\eeq
where $h_n\equiv |d{\rm ln}n_n/dr|_{R_e}^{-1}$ is the scale height of
the neutron distribution at the $\nu_e$--sphere radius ($h_n$ is
actually not really a constant, since the density profile is not
exponential). From this one can estimate the density at the
neutrino--sphere as
\beq
\rho(R_e)\simeq{m_p\over Y_n\sigma h_n}\simeq 5\times 10^{11}{\rm
g\over cm^3}\left({10\ {\rm MeV}\over E_\nu}\right)^2\left({1\ {\rm
km}\over h_n}\right) {\sigma_0\over \sigma},
\eeq
where $Y_n=1-Y_e$ is the neutron fraction ($Y_e\sim 10^{-1}$ and
$h_n\sim 10$~km typically).

Hence, a change in $\sigma$ caused by a large magnetic field will
shift the density of the $\nu_e$--sphere. In particular, for very
large fields $B_*\ga 240(E_\nu/10$~MeV$)^2$, for which only the lowest
Landau level contributes to $\sigma$, one has
$\Sigma=(B_*/2)(m_e/E)^2$, and hence the density of the neutrino
sphere will behave as $B_*^{-1}$ for large fields.

Our results differ significantly from those used in ref.~\cite{bi97},
where the estimate $\Sigma\simeq 0.77 B_*$ (for $B_*\gg 1$) was adopted
from a simple analogy with the $n$--decay results~\cite{ma69}. As a
consequence, to induce a sizeable recoil velocity from the effects on
the neutrino emission resulting from an asymmetric magnetic field
distribution would require much larger fields.

Let us also notice that if $\rho(R_e)$ is bigger than the $\rho_{max}$
discussed in connection with figure~2 (this can happen for small
$E_\nu$), the $\nu_e n\to pe$ cross section will vanish, neglecting
temperature effects, inside the
neutrino sphere. In this case, only magnetic fields large enough to
increase $\rho_{max}$ beyond $\rho(R_e)$ may be able to affect
sizeably the $\nu_e$ opacity.

Considering now the scenario proposed by Kusenko and Segr\`e
\cite{ku96}, we note that a value of $\Sigma>1$ would shift the
$\nu_e$--sphere to larger radii, and hence lower densities. This may
be interesting in order to allow the resonant conversion $\nu_e\to
\nu_\tau$ to happen inside the $\nu_e$--sphere but with smaller values
of $\Delta m^2$ (since $\Delta m^2\propto N_e$), 
possibly within the cosmologically  acceptable values
corresponding to $\sum m_{\nu_i}<92(\Omega_\nu h^2)$~eV. This could cure one of
the main drawbacks of the model, which is the need for large neutrino
masses, with $\Delta m^2>(100\ {\rm eV})^2$. However, for typical
neutrino energies, $E_\nu>5$~MeV, to increase $\Sigma$ significantly
would require quite large fields, $B>10^{16}$~G, and hence this
possibility seems also difficult to implement.

Other mechanisms based on large magnetic fields to produce the
pulsar velocities, such as by means of the $B$ dependence of the
differential cross section of URCA processes \cite{ch84,do85} or of
$\nu e$ scattering \cite{vi95}, do not depend directly on the precise
location of the neutrino spheres, so that the process discussed here 
will not interfere with those scenarios.

In conclusion, we have studied the behaviour of the $\nu_e n\to pe$
cross section in very strong magnetic fields, finding that it
can lead to sizeable modifications of the neutrino opacities in
proto--neutron stars for $B\ga 10^{16}$~G$(E_\nu/10$~MeV$)^2$.
The impact of this for scenarios proposed to explain the
observed pulsar velocities seems then marginal in view of the
extremely high fields required.

\vskip1cm
\acknowledgments{I would like to thank E. Akhmedov, 
D. Grasso, L. B. Leinson, A. Perez and A. Smirnov
for very useful discussions. This work was supported by CICYT, Spain,
under grant No. AEN-96/1718.}

\end{document}